\documentstyle[preprint,prb,aps]{revtex}
\begin{document}
\draft
\title {Charging effects and quantum crossover in granular superconductors}
\author{
E. Granato\\
LAS,
Instituto Nacional de Pesquisas Espaciais,\\
12.225 S\~ao Jos\'e dos Campos, S.P. Brazil. }
\author {
M.A. Continentino\\
Instituto de F\' \i sica,
Universidade Federal Fluminense,\\
Outeiro de S.J.Batista s/n, Niteroi, 24.020, RJ, Brazil. }

\maketitle
\begin{abstract}
The effects of the charging energy in the superconducting  transition of
granular materials or Josephson junction arrays is investigated
using a pseudospin one model.
Within a mean-field
renormalization-group approach, we obtain the
phase diagram as a function of temperature and charging energy.  In contrast to
early treatments, we find no sign of a reentrant transition in agreement with
more recent studies.  A crossover line is identified in the non-superconducting
side of the phase diagram and along which we expect to observe anomalies in the
transport and thermodynamic properties.  We also
study a charge ordering phase, which can appear for large nearest neighbor
Coulomb interaction, and show that it leads to first-order transitions at low
temperatures.  We argue that, in the presence of charge ordering, a non
monotonic behavior with decreasing temperature is possible with a maximum in
the
resistance just before entering the superconducting phase.
\end{abstract}

\section{Introduction}

The recent discovery of high temperature superconductors has renewed the
interest in granular materials. These systems appear to have an intrinsic
"granularity" which is found even in single crystals \cite{deutscher}. A clear
manifestation of this non-uniformity in the scale of the Ginsburg-Landau
coherence length is the two-step nature of the transition to the
superconducting
state  \cite{rosemblatt,borges}. As temperature is lowered, first the
superconducting ordered parameter is developed in each grain at a temperature
$T_{co}$ but  because the thermal energy is higher than the Josephson coupling
between  the grains $E_{J}$, the phases of the order parameter are uncorrelated
due to  thermal fluctuations. Only at a lower temperature $T_c \sim E_J$ will
phase  locking take place leading to long range phase coherence and zero
resistivity. Besides its possible relevance for high temperature
superconductors
granular superconductors have been an active field  of research for many years
\cite{abeles}$^-$\cite{simkin}.

One of the most important issues in the granular superconductors materials is
the role of the charging effects on the phase coherence transition. As pointed
out by Abeles \cite{abeles}, when the grain charging energy $E_c \sim e^2/d$,
where $d$ is the grain diameter and $e$ the electronic charge, is larger than
$E_J$ long-range phase coherence is destroyed due to zero point  fluctuations
of
the phase of the superconducting order parameter. The onset of phase coherence
with increasing intergrain Josephson coupling can then be viewed as a
zero-temperature phase transition. Also, of great interest is the resulting
phase diagram as a function of temperature and charging energy which can
display
reentrant transitions  to the normal state  upon cooling the system to low
temperatures. This possibility and its  experimental observation  have been a
matter of much debate\cite{efetov}$^-$\cite{fazekas84}.
More  recently, an
intriguing effect has been discovered at the onset of superconductivity in
granular systems where the resistivity rises steeply, attaining a sharp
maximum,
before vanishing with decreasing temperature \cite{micklitz}.

In this paper we study a pseudospin one  hamiltonian for granular
superconductors which takes into account self-charging effects in the grains,
the inter-grain ( short-range ) Coulomb interaction and the Josephson coupling
between the phases of the Ginzburg-Landau order parameter of the grains. This
model has been proposed by de Gennes and studied in some detail specially in
relation to the issue of  reentrant behavior \cite{simanek81,fazekas84,fazio}.
Within a mean-field renormalization-group approach, we obtain the phase diagram
as a function of temperature and charging energy but find no sign of a
reentrant
transition in agreement with more recent studies.  We also study the quantum to
classical crossover which takes place in the non-superconducting side of the
phase diagram and identify a crossover line along which we expect to observe
anomalies in the transport and thermodynamic properties with decreasing
temperature.  Finally, we study a charge ordering phase, which can appear for
large nearest neighbor Coulomb interaction \cite{fazekas81}, and show that it
leads to first-order transitions at low temperatures.  Then we argue that, in
the presence of charge ordering, a non monotonic behavior with decreasing
temperature is possible with a maximum in the resistance just before entering
the superconducting phase, which could in principle be observed
experimentally.

\section{Pseudospin One Model}

The standard model for granular superconductors, in the absence of disorder and
dissipation, consists of a regular array of superconducting grains coupled by
Josephson junctions described by the following Hamiltonian
\begin{equation}
H = 2 \sum_{i,j} U_{ij} n_i n_j - E \sum_{<ij>} \cos (\phi_i - \phi_j)
\end{equation}
where $U_{ij}$ is the charging energy due to Coulomb interaction and $E > 0$ is
   the
Josephson energy. Here $n_i$ is the excess of Cooper pairs in the ith grain and
$\phi_i$ is the phase of the order parameter. In the pseudospin one
approximation \cite{simanek81,fazekas84,lebeau} to the above Hamiltonian one
identifies the pair number operator $n_i$ with the $z$-component $S_i^z$ of a
spin one operator. The second term in Eq. (1) when expressed in terms of
$\exp(\pm \phi_i)$ can then be rewritten as raising and lowering operators
$S_i^{\pm}$. If we make the additional approximation of short range Coulomb
interaction this results in  the following Hamiltonian
\begin{equation} H =
2U\sum_{i} [S^z_i]^2
+ 4V\sum_{<i,j>} S^z_iS^z_j - {E\over4}\sum_{<i,j>}(S^+_iS^-_j + S^-_iS^+_j)
\end{equation}
The first term describes intragrain Coulomb energy $U>0$ for different charge
states.   These states, in different grains, are coupled through the
short-range
nearest neighbor Coulomb interaction $V > 0$.  The factors $2$ and $4$ stand
for
the charge of the excitations which are Cooper pairs.  The last term is
ultimately responsible for the phase locking of the different grains.  It is
clear that even at zero temperature this hamiltonian may give rise to phase
transitions due, for example, to a competition between the self-charging term
$U$ and either the intragrain charge interaction $V$ or the Josephson coupling
$E$.   In the former case this competition can lead to an instability with the
formation of charge dipoles for large $V$.  In the latter the competition gives
rise to an off-diagonal long range ordered state characterized by the order
parameter $<S^x>$.  The fact that the above hamiltonian takes into account
only charge fluctuations of $\Delta n = \pm 1 $ is not expected to change the
universality class of the zero-temperature superconductor-insulator transition
considered here.

\section{Mean Field Renormalization Group}

The mean-field renormalization group has been extensivily applied to a variety
of problems both classical and quantum \cite{indekeu,plascak}, with
\cite{mucio}
and without disorder \cite{sergio}.  This method represents an improvement over
the mean-field and the Oguchi pair approximation \cite{fazekas81} since
fluctuations are included at a higher level.  This gives rise for example to
critical exponents which assume non-mean-field values \cite{sergio} and to a
vanishing critical temperature for the two-dimensional Heisemberg model
\cite{plascak}, in agreement with the well known Mermin and Wagner theorem
\cite{mermin}.  We shall employ this method here to investigate the
superconducting transition described by the hamiltonian of Eq.  (2).  The main
reason of doing this is to help settle the issue of the existence (or not) of a
reentrant transition at low temperatures  \cite{efetov}$^-$\cite{fazekas84}.

As mentionned before for large values of $U$ we expect to find the system in a
well defined ( neutral ) charge state and consequently with no phase coherence.
The order parameter describing the phase coherent state, expected to occur when
the Josephson coupling becomes sufficiently large compared to $U$, is the
transverse magnetization $<S^x>$.  The mean-field renormalization group relies
on a scaling relation between quantities calculated using two different finite
systems (or cells) with appropriate boundary conditions.  Within its simplest
version, one consider two cells containing one and two spins each with
corresponding mean fields $b^\prime$ and $b$ acting at the boundaries of the
cells.  For each cell we need to calculate the derivative of the order
parameter
$<S^x>$ for vanishing mean field.  For the spin one model of Eq.  (2) we obtain
for the one spin cell
\begin{displaymath}
{{\partial <S^x>^\prime}
\over{\partial b^\prime}} = {{e\over 2}\, c \,
( {e^{{2\over t^\prime}}} -1) }/ {\,(
2 + {e^{{2\over t^\prime}}} ) }
\end{displaymath}
where we have defined the dimensionless parameters $v=V/U$,  $e=E/U$ and $t=k_B
T/U$ and $c$ is the number of nearest heighbors. Due to the  mean-field like
character of the approach that we discuss here, we expect  the results to be
more likely to hold in three dimensions, i.e., $c=6$.  For the two spin cell we
obtain
\begin{eqnarray*}
{{\partial <S^x>}\over {\partial b}} =
& & {e \over {4
Z_o}}\, \left( c - 1 \right) \,\left(
{{-2\,{e^{{{4\, \left( -1 + v \right) }\over
t}}}}\over {2 - {e\over 2} - 4\,v}} - {2\over {{e^{{{4 + e}\over {2\,t}}}}\,
\left( -2 + {e\over 2} + 4\,v \right) }} \right. \\
& + & {{4\,{e^{{{-4 + e}\over
{2\,t}}}}\, \left( 16 - 32\,e - 5\,{e^2} - 32\,v - 24\,e\,v + 128\,{v^2}
\right)
}\over {64 + 16\,e + 4\,{e^2} + {e^3} + 16\,{e^2}\,v - 256\,{v^2} +
64\,e\,{v^2}}} \\
& - & {4\over {{e^{{{4\,\left( 1 + v \right) }\over t}}}\,
\left( 2 + {e\over 2} + 4\,v \right)}} \\
& + & {{2\, {e^{{{-4 + 4\,v + d}\over
{2\,t}}}} } \over {\left(-e + 4\,v + d \right) }} \\
& \times & \left. { {{\left( -4 -
e + 4\,v - d \right) }^2}\over {\left( {{d^2}\over 2} + 2 d - 2 d v \right) }}
\right)
\end{eqnarray*}
where
\begin{displaymath}
Z_o = {2\over {{e^{{{4 - e}\over
{2\,t}}}}}} + {2\over{{e^{{{4 + e}\over {2\,t}}}}}} + {e^{{{-4\,\left( 1 - v
\right) }\over t}}} + {2\over {{e^{{{4\,\left( 1 + v \right) }\over t}}}}} +
{e^{{{-\left( 4 - 4\,v - d \right) }\over {2\,t}}}} + {e^{{{-\left( 4 - 4\,v +
d
\right) }\over {2\,t}}}}
\end{displaymath}
and $ d = \sqrt{2} \sqrt{8+e^2-16 v + 8 v^2} $.

We now impose the scaling relation \cite{indekeu},
\begin{equation}
{{ \partial <S^x>^\prime } \over {\partial b^\prime}} =
{{\partial <S^x>} \over {\partial b}}
\end{equation}
where the primed quantities are calculated in the smaller cell.  This equation
results from the assumption that the order parameter  and the mean field scale
in the same way and it provides  recursion relations for the couplings or their
ratio $(U/V)$ in terms of the parameters:  temperature, the interaction V and
the number of nearest neighbors $c$ .  The unstable fixed point, as usual in
the
renormalization group procedure, is associated with the critical point at which
the critical exponents are obtained.  The phase boundary obtained from this
recursion relation, for $V/U=0$ and $V/U \ne 0$, is shown in Fig.  2 together
with the result obtained using the Oguchi pair approximation \cite{fazekas81}
which can be regarded here as a self consistent solution of a two site cell
approximation.  As expected, for the same V, the $T=0$ value of the critical
ratio, $(E/U)_c$, for the appearance of the phase-locked state is larger for
the
mean field renormalization group  calculation then for the Oguchi method due to
a better treatment of fluctuations.  Notice that as $V$ increases a larger
value
of $E$ is required to stablish off-diagonal long range order.  Also no sign of
reentrant behavior is found in our results.  We point out that in the
calculation above we assumed no charge ordering which, as will be discussed in
Sec. 5, only holds for \cite{fazekas84} $V/U < v_c$, where $v_c=1/c$.

The
critical line close to the zero temperature fixed point in Fig. 1
rises in temperature as:
\begin{equation}
\left( {E\over U} \right)  - \left( {E\over U} \right)_c \propto \exp (-U/k_B
T)
\end{equation}
The exponential dependence is an artifact of the mean-field nature of the
renormalization group approach and appears for any dimension. In fact for $d=3$
we would expect to find a power law dependence for the critical line, i.e.,
$(E/U)-(E/U)_c  \propto T^{1/\phi }$ where $\phi = \nu z$ is the crossover
exponent. Here, $\nu$ is the correlation-length  and $z$ is the dynamical
critical exponent.  The critical exponents $\nu$ and $z$  are associated with
the  $T=0$ fixed point at $(E/U)_c$. Since this transition is expected to be in
the universality class of the d+1 classical XY model
\cite{doniach81,doniach84}
we have  $z=1$ and $\nu=0.67$ and $1/2$ for $d=2$ and $3$ respectively. We
should point out that our  mean field renormalization group  calculation yields
a lower critical dimension $d_l = 1.95$ for the finite temperature
superconducting instability which is close to the known result for models with
continuous symmetry \cite{mermin}. The existence of a lower critical dimension
within the mean field renormalization group  shows its significant improved
nature compared to the usual mean-field methods.

\section{Quantum Crossover}

As can be seen from the phase diagram in Fig.  1 for $(E/U) < (E/U)_c$
superconductivity in a macroscopic scale is inhibited due to charging effects.
We may however expect to find, even in this non-critical part of the phase
diagram but sufficiently close to the critical point, signs of the incipient
long range superconducting instability.  This should occur as anomalies in the
transport properties, such as  minima in the resistivity  or in thermodynamic
quantities. Where are such anomalies expected to occur?  In a tentative to
clarify this point Fazekas et al.  \cite{fazekas84} and Fazio et al.
\cite{fazio} have calculated the transverse and longitudinal correlation
function respectively for a pair of spins in the non-critical region.  They
found in the phase diagram of Fig. 1 a line of extrema for these quantities
whic
   h
intercept the critical line for $(E/U) > (E/U)_c$ at a finite temperature.
However in their calculations they have completely neglected the effect of
fluctuations of the surrounding medium. Also, if this effect is to be
associated
with a  quantum crossover temperature $T^*$ one would expect from general
scaling arguments, close to the $T=0$ superconductor insulator transiton
\cite{pfeuty,doniach81}, that this temperature should approach zero as $T^*
\sim
({E\over U})_c -{E\over U} )^{z \nu}$.

The fluctuations ignored in the treatment of Ref. \onlinecite{fazekas84,fazio}
c
   an be
taken into account by considering the order parameter "transverse
susceptibility" $\chi^{xx}(T)$ which is related to the "transverse" pair
correlation function averaged over the whole system.  Although this quantity is
not directly related to the magnetic susceptibility it measures the phase
correlations in the whole system \cite{lebeau}.  We have calculated this
quantity as a function of temperature in the non critical region using a two
spin cell approximation and the  result is indicated in Fig.  2.  We note that,
at the critical line $\chi^{xx}(T)$ diverges  as expected.  An important
feature
to be noted is the existence of an inflexion point of the transverse
susceptibility at $T= T^*$ which changes with $E/U$.  The points in which it
occur defines a line in the non-critical region of the phase diagram which is
shown in Fig.  3.  As expected from general scaling arguments \cite{pfeuty}
this
crossover line $T^*(E/U)$ has the same exponential  dependence found for the
critical line for $(E/U) > (E/U)_c$.  The physical meaning of the crossover
line
becomes clear if we recall the competition between charge and phase
fluctuations
which gives rise to the $T=0$ phase transition.  For $T < T^*$ the system is in
a region of rather well defined charge states loosing phase coherence.  As the
temperature is increased for $(E/U) < (E/U)_c$ it enters a regime of strong
charge fluctuations allowing for an enhancement of phase coherence which above
the threshold value $(E/U)_c$ leads to genuine long range phase coherence.
Although the crossover line shown in Fig.  3 suffers from the deficiencies
inherent to mean-field like  calculations there are two points worth stressing:
i) anomalies in physical quantities, like minima in the resistivity, in the
non-critical region of the phase diagram ,i.e., for $(E/U) < (E/U)_c$ , are
expected to occur along the crossover line.  ii) this crossover line is
governed
by the same exponent of the critical line as a consequence of scaling.

The quantum to classical crossover in granular superconductors has also been
studied by Doniach \cite{doniach81}  but we would like to emphasize the
difference between his approach and ours.  While he is studying the crossover
in
the critical region of the phase diagram above the transition line we are
stressing effects which occur for $(E/U) < (E/U)_c$ that is the non-critical
region of the phase diagram.  This region sometimes is more amenable to
experimental observation particularly in the case where critical behavior is
accompanied by extreme critical slowing down which takes the system out of
equilibrium close to the critical line \cite{malozemoff}.  Notice that since by
varying the pressure in a granular material one can in principle alter the
ratio
$(E/U)$ then pressure measurements,
under the assumption that nothing else is changing, would allow
to trace the crossover line by accompanying how anomalies in the physical
quantities shift with applied pressure \cite{borges}.

\section{Charge Ordering }

We have concentrated so far on the effects of phase fluctuations treating the
intergrain coupling V as a small parameter. Let us consider now  how  a larger
V
will affect the superconduting transition.  For this purpose we shall neglect
the Josephson coupling for a while taking $E=0$. Using a spin language it is
clear that while U tries to stablish a singlet (neutral) ground state, the
interaction V favors the existence of local moments ( charge disbalance ) to
take advantage of the lowering of energy due to long range antiferromagnetic
order ( charge instability ). For the system of superconducting grains this
competition gives rise at $T=0$  to a phase transition associated with the
appearance of an insulating  charge ordered state for large V. In fact, the
possibility of an antiferromagnetic ordering of charges has  been considered
sometime ago by Fazekas \cite{fazekas81} who showed that this instability
occurs
for $V/U > v_c$, where $v_c = 1/c$, within  mean field.

To study the effects of $V/U > v_c$ on the phase diagram of Fig. (1) we need to
introduce two order parameters: the transverse magnetization $m_{xy}=<S^x>$,
which describes the  long range phase coherence and the staggered magnetization
$m_z=<S^z>$, which  represents the antiferromagnetic charge ordering. Within a
mean field (one site cell) approximation this amounts to replace Eq. (2) by

\begin{equation}
H= 2 U [S^z] ^2 - 4 V c \, m_z S^z - {E\over 8} c \, m_{xy}\, (S^+ + S^-)
\end{equation}
and the phase boundary is obtained as usual from the solution of
the self consistent equations
\begin{eqnarray}
m_z & = & < S^z > \\ m_{xy}
& = &< S^+ + S^->
\end{eqnarray}

We have solved these equations numerically and obtained a phase diagram as
indicated in Fig.  4.  The important features to be noted are the first order
nature of the superconducting transition at low temperatures, which appear as a
discontinuity in the staggered order parameter $<S^z>$, and the existence of
three thermodynamically different phases.  The first order character of the
superconductor insulator transition at $T=0$, which we find here for large
short
range Coulomb repulsion, i.e.  $V/U > v_c$, is in fact quite similar to the
result of Fisher and Grinstein\cite{mpfisher} for long range Coulomb
interactions where this transition results second or first order depending on
the parameters.  Even more interesting is the topology of the phase diagram of
Fig.  4, with superconducting, charge ordering and normal phases
\cite{footnote}.  If we take into account the temperature dependence of the
Josephson tunneling amplitude $E$ between the grains, which is a decreasing
function of temperature \cite{ambegaoker}, we find that this topology allows
for
the possibility that a granular material in the $V/U > v_c$ regime can become a
charge ordered insulator before becoming superconducting for decreasing
temperature as the system moves through the path indicated in Fig.  4.  Since
in
the charge ordered phase the system should be an insulator, the experimental
signature of this effect could show up as non monotonic behavior of the
resistivity with decreasing temperature which would reach a maximum just before
becoming superconducting.  In fact, an anomalous peak in the resistivity of
some
granular systems has already been observed just before the superconducting
transition \cite{micklitz}.  Note that only for a restricted range of
parameters
near the tricritical point in Fig 4 would this effect be expected which then is
consistent with the rather unusual observation of this phenomena. Of course, a
satisfactory comparison of this result with experiment would require
incorporating into the model several relevant complications as for example
disorder \cite{kosterlitz} and dissipation \cite{chakravarty}.   We expect,
however, that carefully prepared materials could show some signature of this
effect.

\section{Conclusions}

We have studied a pseudospin one model of granular superconductors that takes
into account the competition between charge fluctuations and phase locking in
these materials.  The phase diagram for the phase locking transition has been
obtained within a mean field renormalization group and compared with previous
calculations.  No reentrant behavior has been found within this method in
agreement with recent studies.  We have also shown the existence of a crossover
line in the non-critical region of the phase diagram and along which we expect
to find anomalies in the transport and thermodynamic properties.  This line
could in principle be accessed experimentally by applying external pressure in
the system.  Finally we considered the effects of charge ordering in the phase
diagram within a mean approximation and shown  that it leads to three different
phases and to first order transitions at low temperatures.  We have found that
for a range of intergrain and Josephson coupling interactions the granular
superconductor may enter an insulating charge ordered phase before becoming
superconducting with decreasing temperature.  We suggested that this could lead
to a non monotonic behavior of the resistivity with decreasing temperature with
a maximum in the resistance of the material just before becoming
superconducting.  This could in principle be observed experimentaly as similar
anomalous peaks have already been observed in some granular superconductors.

\section{Acknowledgments}

We thank R. Kishore for helpfull discussons. This work was supported in part by
Conselho Nacional de Desenvolvimento Cient\'ifico e Tecnol\'ogico (CNPq)
(M.A.C.
and E.G.).   M.A.C. also acknowledges the CNPq-RHAE program for supporting a
visit to the Instituto Nacional de Pesquisas Espaciais (INPE) where part of
this
work was done.  He also thanks the members of the research group at INPE for
their kind hospitality.

\begin{figure}
\caption { Critical temperature $T_c$ for the phase-locking transition as
obtained by the mean field renormalization group (MFRG) and a self  consistent
two site cell (Oguchi) approximation. }

\bigskip

\caption { Order parameter transverse susceptibility as function of temperature
for a value of $E/U$ in the insulating phase of Fig. 1. }

\bigskip

\caption { Schematic phase diagram showing the location of the quantum to
classical crossover as determined from the transverse susceptibility.  }

\bigskip

\caption { Phase diagram obtained by mean field approximation for $V/U =0.18$
and $c=6$.  The dot-dashed line indicates first order transitions and the
dashed
line a path described by a sample with a temperature dependent Josephson
coupling $E(T)$. }
\end{figure}

\end{document}